\documentclass[12pt]{iopart}
\usepackage{amssymb,setstack}

\def\ba{\begin{array}}
\def\ea{\end{array}}
\def\be{\begin{equation}}
\def\ee{\end{equation}}
\def\bea{\begin{eqnarray}}
\def\eea{\end{eqnarray}}
\def\nn{\nonumber}
\newtheorem{veta}{Theorem}

\def\pd{\partial}
\def\sp{{\rm span}}

\def\C{{\mathbb C}}
\def\R{{\mathbb R}}
\def\N{{\mathbb N}}

\def\a{{\mathfrak a}}

\def\g{{\mathfrak g}}
\def\h{{\mathfrak h}}

\def\n{{\mathfrak n}}

\def\s{{\mathfrak s}}

\def\t{{\mathfrak t}}
\def\z{{\mathfrak z}}
\def\na{{\mathfrak n}_{n,2}}
\def\DS{{\rm DS}}
\def\CS{{\rm CS}}
\def\US{{\rm US}}
\def\NR{{\rm NR}}

\newcommand{\jpa}{{\it J. Phys. A: Math. Theor.} }

\begin{document}

\title[All solvable extensions of a class of nilpotent Lie algebras]{
All solvable extensions of a class of nilpotent Lie algebras of dimension $n$ and degree of nilpotency $n-1$
}

\author{L \v Snobl\dag\   and P Winternitz\ddag}

\address{\dag\ Faculty of Nuclear Sciences and Physical Engineering, 
Czech Technical University in Prague, B\v rehov\'a 7, 115 19 Prague 1, Czech Republic }

\address{\ddag\ Centre de recherches math\'ematiques and Departement de math\'ematiques et de statistique, 
Universit\'e de Montr\'eal, CP 6128, Succ Centre-Ville, Montr\'eal (Qu\'ebec) H3C 3J7, Canada}

\eads{\mailto{Libor.Snobl@fjfi.cvut.cz}, \mailto{wintern@crm.umontreal.ca}}

\begin{abstract}
We construct all solvable Lie algebras with a specific $n$--dimensional nilradical $\na$ (of degree of nilpotency $n-1$ and with an $n-2$ dimensional maximal Abelian ideal). We find that for given $n$ such a solvable algebra is unique up to isomorphisms. Using the method of moving frames we construct a basis for the Casimir invariants of the nilradical $\na$. We also construct a basis for the generalized Casimir invariants of its solvable extension $\s_{n+1}$ consisting entirely of  rational functions of the chosen invariants of the nilradical.
\end{abstract}



\section{Introduction}\label{intro}

The purpose of this article is to construct all solvable Lie algebras with a  specific nilradical that in an appropriate basis $(e_1,\ldots,e_n)$ has the Lie brackets
\bea
\nn [e_j,e_k] & = & 0, \qquad 1 \leq j, k \leq n-2, \\
\nn [e_{1},e_{n-1}] & = & [e_{2},e_{n-1}] =0, \\
\label{nla} [ e_k,e_{n-1}] & = & e_{k-2}, \qquad 3\leq k \leq n-2,\\
\nn [e_1,e_n] & = & 0,\\
\nn [ e_k,e_n] & = & e_{k-1}, \qquad 2\leq k \leq n-1.
\eea
The nilpotent Lie algebra (\ref{nla}) of dimension $n$ exists for all $n\geq 5$, has degree of nilpotency $n-1$ and has a uniquely defined maximal Abelian ideal $\a$ of dimension $n-2$, equal to its derived algebra.

This article is part of a research program devoted to the classification of Lie algebras over the fields of complex and real numbers. Levi's theorem \cite{Levi,Jac} tells us that any finite dimensional Lie algebra is isomorphic to a semidirect sum of a semisimple Lie algebra and a solvable one. The semisimple Lie algebras have been classified  \cite{Cartan,Gan} (for a more recent reference see \cite{Hel}) and Levi's theorem reduces the classification of all Lie algebras to the classification of solvable ones and some representation theory of the semisimple ones.

Solvable Lie algebras cannot be completely classified. Mubarakzyanov has provided a classification of real and complex Lie algebras of dimension $n\leq 5$ \cite{Mub1,Mub2} (equivalent in dimension 3 to previous classifications by Bianchi \cite{Bianchi} and Lie \cite{Lie}) and a partial classification for $n=6$ \cite{Mub3}. 
The classification for $n=6$ was continued by Turkowski \cite{Tur2} who also considered the classification of semidirect sums of semisimple and solvable Lie algebras \cite{Tur1,Tur3}. 

The method of classifying and constructing solvable Lie algebras used by Mubarakzyanov and Turkowski was based on the fact that every solvable Lie algebra has
a uniquely defined nilradical $\n = \NR ( \s)$ \cite{Jac}. Its dimension satisfies \cite{Mub1}
\be\label{dimnilrad}
{\rm dim} \ \n \geq \frac{1}{2} \ {\rm dim} \ \s.
\ee
Hence we can consider a given nilpotent Lie algebra $\n$ of dimension $n$ and classify all of its extensions to solvable Lie algebras. This is an open ended task since infinitely many different series of nilpotent Lie algebras exist and they themselves have not been classified.

Nilpotent Lie algebras of dimension $n=6$ over complex numbers were classified by Umlauf \cite{Umlauf}, over an arbitrary field of characteristic zero by Morozov \cite{Morozov} who also gives a list of all lower dimensional ones. Those of dimension $n=7$ and $n=8$ were classified by Safiulina \cite{Saf} and Tsagas \cite{Tsa1}, respectively, and some results for $n=9$ are known \cite{Tsa2}. We mention that the number of nonequivalent nilpotent algebras increases very rapidly with their dimension $n$ and for $n\geq 7$ becomes infinite, i.e. classes of nonisomorphic nilpotent algebras depending on parameters arise. For reviews of this field of research with extensive bibliographies see e.g. \cite{GKh,Kh}. For classifications of specific type of nilpotent Lie algebras with $n\leq 9$ see e.g. \cite{AC,ACS,GKh,Kh}.

A more manageable task is to start from series of nilpotent Lie algebras that already exist in low dimensions.
Thus, for $n=1,2$ a nilpotent Lie algebra must be Abelian. For $n=2k+1,k\geq 1$ another series exists, namely the Heisenberg algebras. A further series exists for all $n\geq 4$. For lack of better name it was called $\n_{n,1}$ in our earlier article \cite{SW}. Six inequivalent indecomposable nilpotent Lie algebras exist for $n=5$, among them a Heisenberg algebra, the algebra $\n_{5,1}$ and also the algebra which we shall denote $\n_{5,2}$ (and more generally $\n_{n,2}$, $n\geq 5$) with the commutation relations as in Eq. (\ref{nla}) above.

Earlier articles were devoted to solvable extensions of Heisenberg algebras \cite{RW}, Abelian Lie algebras \cite{NW,NW1}, ``triangular'' Lie algebras \cite{TW,TW1} and the algebras $\n_{n,1}$ \cite{SW}.

The motivation for providing a classification of Lie algebras was discussed in our previous article \cite{SW}. Let us mention that string theory and other elementary particle theories require the use of higher dimensional spaces. A classification of such spaces, analogous to the Petrov classification of Einstein spaces  \cite{Petrov} is based on 
the classification of higher dimensional Lie groups, in particular the solvable ones.

A more general reason why a classification of Lie algebras (and by extension Lie groups) is needed in physics is that Lie groups typically occur as groups of transformations of the solution space of some equations. These equations may be of differential, difference, integral or some other type. Different systems may have isomorphic symmetry groups and in this case the results obtained for one theory can be transferred to another one. 

In the representation theory of Lie algebras and Lie groups an important role is played by Casimir operators or generalized Casimir operators, i.e. polynomial and nonpolynomial invariants of the coadjoint representation. Casimir invariants, corresponding to elements of the center of the enveloping algebra of a Lie algebra are of primordial importance in physics. They represent such important quantities as angular momentum, elementary particle mass and spin, Hamiltonians of various physical systems etc.
Also the generalized Casimir invariants occur in physics. Indeed, Hamiltonians and integrals of motion for classical integrable Hamiltonian systems are not necessarily polynomials in the momenta \cite{Hietarinta,Ramani}, though typically they are invariants of some group action. 

Part of our classification program \cite{NW,NW1,RW,SW,TW,TW1} is to construct a basis, i.e. a maximal set of functionally independent generalized Casimir operators of each algebra obtained in the classification. We do this for the algebras $\na$ below in Section \ref{GCI}.

The present article is organized as follows. Section \ref{mathprelim} is devoted to some mathematical preliminaries. In Section \ref{cslawn} we present the complete classification of solvable Lie algebras with the nilradical $\na$ and show that precisely one non--nilpotent element can be added. This element has a diagonal action on the nilradical. In Section \ref{GCI} we first calculate the Casimir invariants of the nilpotent algebras $\na$, i.e. the polynomial invariants of the coadjoint representation of $\na$. There exist $n-4$ functionally independent Casimir invariants $\xi_0=e_1,\xi_1,\ldots,\xi_{n-5}$. We then show that the solvable Lie algebra $\s_{n+1}$ has exactly $n-5$ functionally independent generalized Casimir operators which can be chosen in the form
\be
\chi_j = \frac{\xi_j^{n}}{\xi_0^{(n-1)(j+2)}}, \ 1 \leq j \leq n-5.
\ee

\section{Mathematical preliminaries}\label{mathprelim}

\subsection{Basic concepts}

Three different series of subalgebras can be associated with any given Lie algebra. The dimensions of the subalgebras in each of these series are important characteristics of the given Lie algebra.

The {\it derived series} 
$ \g = \g^{(0)} \supseteq \g^{(1)} \supseteq \ldots \supseteq \g^{(k)} \supseteq \ldots $ 
is defined recursively
\begin{equation}
 \g^{(0)}=\g, \qquad  \g^{(k)} = [\g^{(k-1)},\g^{(k-1)}], \ k\geq 1.
\end{equation}
If the derived series terminates, i.e. there exists $k \in \N$ such that $\g^{(k)} = 0$, then $\g$ is called a {\it solvable Lie algebra}.

The {\it lower central series} $ \g = \g^{1} \supseteq \g^{2} \supseteq \ldots \supseteq \g^{k} \supseteq \ldots $ 
is again defined recursively
\begin{equation}
\g^{1}=\g, \qquad \g^{k} = [\g^{k-1},\g], \ k\geq 2. 
\end{equation}
If the lower central series terminates, i.e. there exists $k \in \N$ such that $\g^{k} = 0$, then $\g$ is 
called a {\it nilpotent Lie algebra}. The highest value of $k$ for which we have $\g^{k} \neq 0$ is the degree of 
nilpotency of a nilpotent Lie algebra.

Obviously, a nilpotent Lie algebra is also solvable. An Abelian Lie algebra is nilpotent of degree 1.

The {\it upper central series} is $ \z_{1} \subseteq \ldots \subseteq \z_{k} \subseteq \ldots \subseteq \g$. 
In this series $\z_1$ is the {\it center} of $\g$
\begin{equation}
\z_1 = C(\g) = \{ x \in \g | [x,y]=0, \ \forall y \in \g \}.
\end{equation}
Further we define recursively $\z_k$ as the unique ideal in $\g$ such that 
$\z_k/\z_{k-1}$ is the center of $\g/\z_{k-1}$. The upper central series terminates, i.e. a number $k$ exists such that $\z_k=\g$, if and only if $\g$ is nilpotent \cite{Jac}.

We shall call these three series the {\it characteristic series} of the algebra $\g$. We shall use the notations $\DS,\CS$ and $\US$ for (ordered) lists of integers denoting the dimensions of subalgebras in the derived, lower central and upper central series, respectively. We list the last (then repeated) entry only once (e.g. we write $\CS=[n,n-1]$ rather than $\CS=[n,n-1,n-1,n-1,\ldots]$).

The {\it centralizer} $\g_\h$ of a given subalgebra $\h \subset \g$ in $\g$ is the set of all elements in $\g$ commuting with all
elements in $\h$, i.e.
\be
\g_\h = \{ x \in \g | [x,y]=0, \ \forall y \in \h \} .
\ee

An {\it automorphism} $\Phi$ of a given Lie algebra $\g$ is a bijective linear map
\[\Phi: \ \g \rightarrow \g \]
such that for any pair $x,y$ of elements of $\g$ 
\be\label{autom} 
\Phi([x,y])=[\Phi(x),\Phi(y)]. 
\ee
We recall that all automorphisms of $\g$ form a Lie group ${\rm Aut}(\g)$.
Its Lie algebra is then the algebra of {\it derivations} of $\g$, i.e. of linear maps 
\[D: \ \g \rightarrow \g \]
such that for any pair $x,y$ of elements of $\g$ 
\be\label{deriv} 
D([x,y])=[D(x),y]+[x,D(y)]. 
\ee
If an element $z \in \g$ exists,  such that 
\[ D = {\rm ad} (z), \; \; {\rm i.e.} \; D(x)=[z,x], \ \forall x \in G,  \] 
the derivation is called an {\it inner derivation}, any other one is an {\it outer derivation}.

\subsection{Solvable Lie algebras with a given nilradical}\label{introsan}

Any solvable Lie algebra $\s$ contains a unique maximal nilpotent ideal
$\n={\NR}(\s)$, the {\it nilradical} $\n$.
We will assume that $\n$ is known. That is, in some basis $( e_1, \ldots, e_n )$ of $\n$ we know the Lie brackets
\be\label{nilkom}
[e_j,e_k] = N^l_{jk} e_l
\ee
(summation over repeated indices applies).
We wish to extend the nilpotent algebra $\n$ to all possible indecomposable solvable Lie algebras $\s$ having $\n$ as their
nilradical. Thus, we add further elements $f_1,\ldots,f_f$ to the basis $( e_1, \ldots, e_n )$ which together 
form a basis of $\s$. The derived algebra of a solvable Lie algebra is contained in the nilradical (see \cite{Jac}), i.e.
\be\label{ssinn}
[\s,\s] \subseteq \n.
\ee
It follows that the Lie brackets on $\s$ take the form
\bea\label{Agam1}
[f_a,e_j] & = & (A_a)^k_{j} e_k, \; 1 \leq a \leq f, \  1 \leq j \leq n, \\ 
\label{Agam2} [f_a,f_b] & = & \gamma^j_{ab} e_j, \; 1 \leq a,b \leq f.
\eea

The matrix elements of the matrices $A_a$ must satisfy certain linear relations following from the Jacobi relations between the elements $(f_a,e_j,e_k)$. The Jacobi identities between the triples  $(f_a,f_b,e_j)$ will provide 
linear expressions for the structure constants $\gamma^j_{ab}$ in terms of the matrix elements of the commutators of the matrices $A_a$ and $A_b$.

Since $\n$ is the maximal nilpotent ideal of $\s$ no nontrivial 
linear combination of the matrices $A_i$  is a nilpotent matrix, i.e. they are {\it linearly nil--independent}.

Let us now consider the adjoint representation of $\s$, restrict it to the nilradical $\n$ and find 
${\rm ad}|_\n (f_a)$. 
It follows from the Jacobi identities that ${\rm ad}|_\n (f_a)$ is 
a derivation of $\n$. In other words, finding all sets of matrices $A_a$ in (\ref{Agam1}) satisfying the Jacobi identities
is equivalent to finding all sets of outer nil--independent derivations of $\n$
\be
 D^1={\rm ad}|_\n (f_1),\ldots,D^f={\rm ad}|_\n (f_f).
\ee
Furthermore, in view of (\ref{ssinn}), the commutators $[D^a,D^b]$ must be inner derivations of $\n$. This requirement 
determines the Lie brackets (\ref{Agam2}), i.e.  the structure constants $\gamma^j_{ab}$, up to elements in the center $C(\n)$ of $\n$.

Different sets of derivations may correspond to isomorphic Lie algebras, so redundancies must be eliminated. 
The equivalence is generated by the following transformations:
\begin{enumerate}
\item We may add any inner derivation to $D^a$.
\item We may perform a change of basis in $\n$ such that the Lie brackets (\ref{nilkom}) are not changed.
\item We can change the basis in the space $ \sp \{ D^1,\ldots, D^f \}$.
\end{enumerate}

\section{Classification of solvable Lie algebras with the nilradical $\na$}\label{cslawn}

\subsection{Nilpotent algebra $\na$ and its structure}

The Lie algebra $\n=\na$ is defined by the Lie brackets (\ref{nla}) of the Introduction.
We shall mostly consider $n \geq 6$ (the final result is the same for $n=5$ but there is a small peculiarity in the computation).
The dimensions of the subalgebras in the characteristic series are
\be
 \DS=[n,n-2,0], \; \CS=[n,n-2,n-3,\ldots,1,0], \; \US=[1,2,\ldots,n-2,n].
\ee
Its maximal Abelian ideal $\a$ coincides with the derived algebra $\n^{(1)}=\n^2$, i.e. $\a =\sp \{ e_1, \ldots,e_{n-2} \}$. 

In order to find all non--nilpotent derivations of $\n$ we first consider the structure of automorphisms of $\na$.
There exists a flag of ideals which is invariant under any automorphism
\be\label{flag}
\n\supset \n_{\n^{n-2}}\supset \n^2 \supset \n^3 \supset \ldots \supset \n^{n-1}
\ee
where each element in the flag has codimension one in the previous one. (We recall that $\n_{\n^{n-2}}$ is the centralizer of $\n^{n-2}$ in $\n$.) In any basis respecting the flag, e.g. the one used in the Lie brackets (\ref{nla}), any automorphism will be represented by a triangular matrix.

Furthermore, the whole algebra $\n$ is generated via multiple commutators of the elements $e_{n-1}$ and $e_n$, e.g.
$e_{n-3}=[[e_{n-1},e_{n}],e_n]$. That means that due to the definition of an automorphism (\ref{autom}) 
the knowledge of
\be\label{defgen}
\Phi(e_{n-1})=\sum_{k=1}^{n-1} \phi_k e_k, \qquad \Phi(e_{n})=\sum_{k=1}^{n} \psi_k e_k
\ee
in principle amounts to full knowledge of $\Phi$. It remains to establish which choices of $\phi_k,\, 1\leq k\leq n-1$ and $\psi_k,\, 1\leq k\leq n$ are consistent with the definition (\ref{autom}) of an automorphism.

Because of the triangular structure respecting (\ref{flag}) we immediately have
\bea
\nn [\Phi(e_j),\Phi(e_k)] & = & 0, \; 1 \leq j, k \leq n-2, \\
\nn [\Phi(e_{1}),\Phi(e_{n-1})] & = & [\Phi(e_{2}),\Phi(e_{n-1})] =0, \\
\nn [\Phi(e_1),\Phi(e_n)] & = & 0
\eea
and relation
\be
[\Phi(e_k),\Phi(e_n)]  =  \Phi(e_{k-1}), \; 2\leq k \leq n-1
\ee
can be viewed as a definition of $\Phi(e_{k-1}), \; 2\leq k \leq n-1$ in accordance with (\ref{defgen}). 
Consequently it remains to check
$$ [ \Phi(e_k),\Phi(e_{n-1})] =  \Phi(e_{k-2}), \; 3\leq k \leq n-2 $$
or, equivalently,
\be\label{restriction}
[ \Phi(e_{n-1}),\Phi(e_k)]  = - [ \Phi(e_{n}),[\Phi(e_n),\Phi(e_k)]], \; 1\leq k \leq n-2
\ee
(the change in the index range is just for convenience, the added two relations are satisfied trivially).
Since any automorphism $\Phi$ restricted to $\n^2={\rm span} \{e_1,\ldots,e_{n-2} \}$ is a regular (invertible) map, we can write (\ref{restriction}) as a relation between restrictions to $\n^2$ of adjoint operators
\be\label{restriction1}
{\rm ad}|_{\n^2} (\Phi(e_{n-1})) = - \left( {\rm ad}|_{\n^2} (\Phi(e_{n})) \right)^2.
\ee
Writing down the matrices of the operators we find
$$ {\rm ad}|_{\n^2} (\Phi(e_{n-1})) = \left( \ba{cccccccc}
0 & \, 0 & -\phi_{n-1} &  0 & 0 & \ldots & 0 \\
 & \, 0 & 0 & -\phi_{n-1} & 0   & \ldots & 0 \\
 &   & 0 & 0 & -\phi_{n-1}   & \ldots & 0 \\
 &  & & \ddots & \ddots & \ddots &  \\
 & &  &        & 0      & 0 & -\phi_{n-1} \\
 & &  &        &        & 0 & 0 \\
& & &         &        &   & 0 \ea \right) $$
and
$$ {\rm ad}|_{\n^2} (\Phi(e_{n})) = \left( \ba{cccccccc}
0 & -\psi_n & -\psi_{n-1} &  0 & 0 & \ldots & 0 \\
 & 0 & -\psi_n & -\psi_{n-1} & 0   & \ldots & 0 \\
 &   & 0 & -\psi_n & -\psi_{n-1}   & \ldots & 0 \\
 &  & & \ddots & \ddots & \ddots &  \\
 & &  &        & 0      & -\psi_n & -\psi_{n-1} \\
 & &  &        &        & 0 & -\psi_n \\
& & &         &        &   & 0 \ea \right). $$
Consequently, the condition (\ref{restriction1}) gives us two constraints on $\phi_k,\, \psi_k$, namely
\be\label{restriction2}
\phi_{n-1}=\psi_{n}^2, \qquad \psi_{n-1}=0.
\ee
Note that here the case $n=5$ differs -- the matrices above become $3\times3$ and only the first condition remains. Accordingly, the dimension of the group of automorphisms and the algebra of derivations is by one higher than in the generic case. Even so, the final conclusion about the number and structure of the solvable Lie algebras with the nilradical $\n_{5,2}$ also fits into the general pattern shown below.

To sum up, we have found that automorphisms $\Phi$ of $\n$ are uniquely determined by a set of $2n-3$ parameters
\be
\phi_k,\psi_k,\psi_n \qquad  1\leq k\leq n-2
\ee
where $\psi_n\neq 0$ and $\phi_k,\psi_k$ are arbitrary.
Such an automorphism acts on the basis elements $e_k$ in the following way
\bea
\nonumber \Phi(e_k) & = & \sum_{j=1}^{k-1} (\ldots) \, e_j+ (\psi_n)^{n-k+1} \, e_k, \qquad 1\leq k \leq n-2, \\
\Phi(e_{n-1}) & = & \sum_{j=1}^{n-2} \phi_j \, e_j+\psi_n^2 \, e_{n-1}, \\
\nonumber \Phi(e_{n}) & = & \sum_{j=1}^{n-2} \psi_j \, e_j+\psi_n \, e_{n}
\eea
where the coefficient of the $e_k$ term in $\Phi(e_k)$ was found from (\ref{autom}) and $\ldots$ denote some rather complicated functions of the parameters $\phi_k,\psi_k,\psi_n$
which can be deduced from Eq. (\ref{autom}) but we shall not need them in the following.

The derivations of $\n$ are now easily found by considering automorphisms infinitesimally close to identity, i.e. differentiating one--parameter subgroups in ${\rm Aut}(\n)$. We find that the algebra of derivations is $2n-3$ dimensional. An arbitrary derivation $D$ depends on $2n-3$ parameters $c_k,d_k,d_n,\; 1\leq k\leq n-2$ and has the form
\bea
\nonumber D(e_k) & = & \sum_{j=1}^{k-1} (\ldots) \, e_j+ (n-k+1) \, d_n \, e_k, \qquad 1\leq k \leq n-2, \\
D(e_{n-1}) & = & \sum_{j=1}^{n-2} c_j \, e_j+2 \, d_n \, e_{n-1}, \label{genderiv} \\
\nonumber D(e_{n}) & = & \sum_{j=1}^{n-2} d_j \, e_j+d_n \, e_{n}
\eea
where $\ldots$ denote some linear functions of the parameters $c_k,d_k,d_n$ 
(again their explicit knowledge is not needed in the rest of the paper).

\subsection{Construction of solvable Lie algebras with nilradical $\na$}\label{csla}

As was explained in Subsection \ref{introsan}, to find all solvable Lie algebras with nilradical $\na$  we must find all nonequivalent nil--independent sets $ \{ D^1,\ldots, D^f \} $ of derivations $\na$. 

Looking at (\ref{genderiv}) we immediately recognize that we can have at most one nil--independent derivation -- such that $d_n \neq 0$. If there would be more of them, say $D$ and $\tilde D$ then obviously by taking a linear combination $\tilde{d}_n D-d_n \tilde{D}$ we obtain a nilpotent operator (namely one represented by a strictly upper triangular matrix). Therefore any solvable but not nilpotent Lie algebra with the nilradical $\na$ must be $n+1$ dimensional. The question which remains is how many such algebras are non--isomorphic. 

By proper choice of the multiple of $D$ and adding suitable inner derivations we can transform $D$ into the form
\bea
\nonumber D(e_k) & = & \sum_{j=1}^{k-1} (\ldots) \, e_j+ (n-k+1)  \, e_k, \qquad 1\leq k \leq n-2, \\
D(e_{n-1}) & = & \sum_{j=1}^{n-3} c_j \, e_j+2 \, e_{n-1}, \label{specderiv} \\
\nonumber D(e_{n}) & = &  e_{n}.
\eea
There are $n-1$ nontrivial inner derivations ${\rm ad} (e_k), \; 2\leq k \leq n$ and one choice of scaling, so we 
are able to remove $n$ parameters in a non--nilpotent outer derivation (\ref{genderiv}). There are still $n-3$ parameters left in Eq. (\ref{specderiv}).

Next we perform a change of basis in $\n$ such that the Lie brackets (\ref{nla}) are preserved, i.e. conjugate the derivation $D$ by a suitable automorphism $\Phi$
\begin{equation}
 D \rightarrow \tilde{D}=\Phi^{-1} \circ D \circ \Phi. 
\end{equation}
Our aim is to diagonalize the action of $D$, if possible. We find it convenient to perform this in $n-3$ steps, setting one parameter $c_k$ equal to $0$ in each step. Thus our $\Phi$ will be expressed as 
\begin{equation}\label{phiprod}
 \Phi = \Phi_{n-3} \circ \Phi_{n-2} \circ \cdots \circ \Phi_{1}
\end{equation}
where the automorphisms $\Phi_k$ are constructed as follows.

Let us assume that for a given $k\leq n-3$ we have already set $c_j=0$ for all $k<j\leq n-2$ (assuming of course the form (\ref{specderiv}) for $D$). We construct an automorphism $\Phi_k$ defined by
$$\Phi_k(e_{n-1}) = \alpha_k e_{k}+e_{n-1}, \qquad \Phi_k(e_{n}) = e_n$$
where $\alpha_k$ is to be determined. We have
$$ D(\Phi_k(e_{n-1}) )= D(e_{n-1}) + \alpha_k D(e_k) = 2 e_{n-1} + c_k e_k + (n-k+1) \alpha_k e_k+\sum_{j=1}^{k-1} (\ldots) e_j $$
$$ = 2 \left( e_{n-1} +\frac{1}{2} \left( c_k+(n-k+1) \alpha_k \right) e_k\right) + \sum_{j=1}^{k-1} (\ldots) e_j.$$
We find that 
$$ D(\Phi_k(e_{n-1}) )= 2 \, \Phi_k(e_{n-1})  + \sum_{j=1}^{k-1} (\ldots) \, e_j$$
precisely when 
$$ \alpha_k = \frac{c_k}{k-n+1}. $$
By this choice of $\alpha_k$ we set $c_k$ to $0$ and proceed to the next step, namely elimination of $c_{k-1}$.

To conclude, we are able to eliminate all $c_k$'s using suitably chosen automorphisms $\Phi_k$ in Eq. (\ref{phiprod}), i.e. we have found that up to addition of inner derivations, conjugation by automorphisms, and rescaling there exists just one nil--independent set of outer derivations, consisting of a unique element $D$
\be\label{uniquederiv}
D(e_k)  =  (n-k+1)  \, e_k, \qquad 1\leq k \leq n.
\ee
Consequently, we have

\begin{veta}\label{th1}
For the given nilradical $\na$, $n\geq 5$ there exists precisely one solvable non--nilpotent Lie algebra $\s_{n+1}$ with the nilradical $\na$. It has dimension $\dim \s_{n+1}=n+1$ and its Lie brackets are as follows
\bea
\nn  & \s_{n+1} = & {\rm span} \{ e_1,\ldots,e_k,f_1 \}, \\
\nn [e_j,e_k] & = & 0, \qquad 1 \leq j, k \leq n-2, \\
\nn [e_{1},e_{n-1}] & = & [e_{2},e_{n-1}] =0, \\
\label{sla} [ e_k,e_{n-1}] & = & e_{k-2}, \qquad 3\leq k \leq n-2,\\
\nn [e_1,e_n] & = & 0,\\
\nn [ e_k,e_n] & = & e_{k-1}, \qquad 2\leq k \leq n-1, \\
\nn [ e_k,f_1] & = & (n-k+1) e_{k}, \qquad 1 \leq k \leq n.
\eea
The dimensions of the characteristic series are
\bea
\nn \DS  & = &  [n+1,n,n-2,0], \ \CS=[n+1,n], \ \US=[0]. 
\eea
\end{veta}
Above we have proved Theorem \ref{th1} for $n\geq 6$. For $n=5$ the proof requires a slight modification (see a comment below Eq. (\ref{restriction2})) but proceeds in a very similar way. One just has to construct one more automorphism eliminating an additional parameter in the non--nilpotent derivation $D$.

\section{Generalized Casimir invariants}\label{GCI}

\subsection{Definitions and methods of computation}

The term {\it Casimir operator}, or {\it Casimir invariant}, is usually reserved for elements of the center of the enveloping algebra of a Lie algebra $\g$ \cite{Casimir,Racah}. These operators are in one--to--one correspondence with polynomial invariants characterizing orbits of the coadjoint representation of $\g$ \cite{Kirillov} (or of the corresponding Lie group $G$). On the other hand, in the representation theory of solvable Lie algebras the invariants of the coadjoint representation are not necessarily polynomials. They can be rational functions, or even transcendental ones. In that case we call them {\it generalized Casimir invariants}. For algebraic Lie algebras, which include semisimple, perfect and also nilpotent Lie algebras it is possible to choose a basis for all invariants of the coadjoint representation consisting entirely of polynomials \cite{Abellanas, Abellanas1}.

Two different systematic methods of constructing invariants of group actions exist, in particular of the coadjoint representation of a Lie group $G$. The first method is an infinitesimal one. A basis for the coadjoint representation of the Lie algebra $\g$ of the Lie group $G$ is given by the first order differential operators
\be\label{doal}
 \hat X_k = x_a c^a_{kb} \frac{\pd}{\pd x_b}.
\ee
where $c^k_{ij}$ are the structure constants of Lie algebra $\g$ in the basis $(x_1,\ldots,x_N)$.
In Eq. (\ref{doal}) the quantities $x_a$ are commuting independent variables which can be identified with coordinates in the basis of the space $\g^*$ dual to the basis $(x_1,\ldots,x_N)$ of the algebra $\g$.

The invariants of the coadjoint representation, i.e. the generalized Casimir invariants, are solutions of the following system of partial differential equations
\be\label{casimir}
 \hat X_k I(x_1, \ldots,x_N)=0, \ k=1,\ldots,N .
\ee
Traditionally, the system (\ref{casimir}) is solved using the method of characteristics.

The number of functionally independent solutions of the system (\ref{casimir}) is
\be\label{nomcas}
n_I=N-r
\ee
where $r$ is the generic rank of the antisymmetric matrix
\be\label{nomcas2}
C = \left( \ba{cccc} 0 & c^b_{12} x_b & \ldots & c^b_{1N} x_b \\
-c^b_{12} x_b & 0 & \ldots & c^b_{2N} x_b \\
\vdots & & & \vdots \\
-c^b_{1,N-1} x_b & \ldots & 0 & c^b_{N-1,N} x_b \\
-c^b_{1N} x_b & \ldots & -c^b_{N-1,N} x_b & 0 \ea \right).
\ee
Since $C$ is antisymmetric, its rank is even. 
Hence $n_I$ has the same parity as $N$. 

This method of calculating the invariants of Lie group actions is a standard one and goes back to the 19th century. For a brief history with references to the original literature we refer to Olver's book \cite{Olver}. To our knowledge this method was first adapted to the construction of (generalized) Casimir operators in \cite{Abellanas,Abellanas1,BB}. It has been extensively applied to low dimensional Lie algebras (for $n\leq 5$ and nilpotent $n=6$ in \cite{PSW}, solvable $n=6$ in \cite{Ndo,BPP1} with 4--dimensional nilradicals and in \cite{CS4} with 5--dimensional nilradicals), certain solvable rigid Lie algebras \cite{CS,CS1}, solvable Lie algebras with Heisenberg nilradical \cite{RW}, the nilradicals ${\mathfrak n}_{n,1}$ \cite{SW}, triangular nilradicals \cite{TW1}, certain inhomogeneous classical Lie algebras \cite{CS3}, certain affine Lie algebras \cite{CS2} and other specific solvable Lie algebras
\cite{AC,ACS}.

The second method of calculating invariants of group actions is called the method of moving frames. It goes back to Cartan \cite{CartanMF1,CartanMF2} and its recent formulation is due to M. Fels and P. Olver \cite{FelsOlver1,FelsOlver2}. A related method was also applied to the inhomogeneous classical groups \cite{Perroud}). Boyko et al. adapted the method of moving frames to the case of coadjoint representations. They presented an algebraic algorithm for calculating (generalized) Casimir operators and applied it to a large number of solvable Lie algebras \cite{BPP1,BPP2,BPP3,BPP4,BPP5}.

We shall apply the method of moving frames to calculate the invariants of the coadjoint action of the groups corresponding to the nilpotent Lie algebras $\na$ and the solvable Lie algebra $\s_{n+1}$ of Theorem \ref{th1}.

The method of moving frames as we apply it can be roughly divided into the following steps.
\begin{enumerate}
\item Integration of the coadjoint action of the Lie algebra $\g$ on its dual $\g^*$ as given by the vector fields (\ref{doal}) to the (local) action of the group $G$.

This is usually realized by choosing a convenient (local) parameterization of $G$ in terms of 1--parametric subgroups, e.g.
\begin{equation}\label{localgrparam}
g(\vec \alpha) = \exp(\alpha_N x_N)\cdot \ldots \cdot \exp(\alpha_2 x_2)\cdot \exp(\alpha_1 x_1) \in G, \quad \vec \alpha = (\alpha_1,\ldots,\alpha_N)
\end{equation}
and correspondingly composing the flows $\Psi^{\alpha_k}_{\hat X_k}$ of the vector fields $\hat X_k$ defined in (\ref{doal})
\begin{equation}
\frac{{\rm d}\Psi^{\alpha_k}_{\hat X_k}(p)}{{\rm d \alpha_k}} = \hat X_k(\Psi^{\alpha_k}_{\hat X_k}(p)),\qquad p\in\g^*,
\end{equation}
i.e. we have
\begin{equation}
\Psi(g(\vec \alpha))= \Psi^{\alpha_N}_{\hat X_N} \circ \ldots \Psi^{\alpha_2}_{\hat X_2} \circ \Psi^{\alpha_1}_{\hat X_1}.
\end{equation}
For a given point $p\in\g^*$ with coordinates $x_k=x_k(p), \; \vec x=(x_1,\ldots,x_N)$
we denote the coordinates of the transformed point $\Psi(g(\vec \alpha))p$ by $\tilde{x}_k$
\begin{equation}\label{transfx}
\tilde{x}_k \equiv \Psi_k (\vec \alpha) (\vec x) = x_k\left( \Psi(g(\vec \alpha))p \right).
\end{equation}
We consider $\tilde{x}_k$ to be a function of both the group parameters $\vec \alpha$ and the coordinates $\vec x$ of the original point $p$. 

\item Choice of a section cutting through the orbits of the action $\Psi$.

We need to choose in a smooth way a single point on each of the (generic) orbits of the action of the group $G$. Typically this is done as follows: we find a subset of $r$ coordinates, say $(x_{\pi(i)})_{i=1}^{r}$, on which the group $G$ acts transitively, at least locally in an open neighborhood of chosen values $(x^0_{\pi(i)})_{i=1}^{r}$.  Here $\pi$ denotes a suitable injection $\pi:\{1,\ldots,r\}\rightarrow \{ 1,\ldots,N \}$ and $r$ is the rank of the matrix $C$ in Eq. (\ref{nomcas2}). Points whose coordinates satisfy 
\begin{equation}\label{section}
x_{\pi(i)}=x^0_{\pi(i)}, \; 1 \leq i \leq r
\end{equation}
form our desired section $\Sigma$, intersecting each generic orbit once.
\item Construction of invariants.

For a given point $p\in\g^*$ we find group elements transforming $p$ into $\tilde{p}\in\Sigma$ by the action $\Psi$. We express as many of their parameters as possible (i.e. $r$ of them) in terms of the original coordinates 
$\vec x$ and substitute them back into Eq. (\ref{transfx}). This gives us $\tilde{x}_k$ as functions of $\vec x$ only. Out of them, $\tilde{x}_{\pi(i)},\; i=1,\ldots,r$ have the prescribed fixed values $x^0_{\pi(i)}$. The remaining $N-r$ functions $\tilde{x}_k$ are by construction invariant under the coadjoint action of $G$, i.e. define the sought after invariants of the coadjoint representation.
\end{enumerate}

Technically, as we shall see in our particular case below (cf. Eq. (\ref{tildee}), it may not be necessary to evaluate all the functions $\tilde{x}_k$ so that a suitable choice of the basis in $\g$ can substantially simplify the whole procedure. This happens when only a smaller subset of say $r_0$ group parameters $\alpha_k$ enters into the computation of $N-r+r_0$ functions $\tilde{x}_k, \; k=1,\ldots,N-r+r_0$ (possibly after a re--arrangement of $x_k$'s). In this case the other parameters can be ignored throughout the computation. They are specified by the remaining equations 
\begin{equation}
\tilde{x}_{i}=x^0_{i}, \; N-r+r_0+1 \leq i \leq N
\end{equation}
but do not enter into the expressions for $\tilde{x}_k, \; 1 \leq k \leq N-r+r_0$ which define our invariants.

We shall remark that the invariants found using either of the methods above may not be in the most 
convenient form. That can be remedied once we find them. For example, as we already mentioned, the generalized Casimir invariants of a nilpotent Lie algebra can be always chosen as polynomials, i.e. proper Casimir invariants. As we shall see below the method of moving frames may naturally give us nonpolynomial ones. Nevertheless, it is usually quite easy to construct polynomials out of them.

\subsection{Casimir invariants of the Lie algebra $\na$}

The differential operators corresponding to the basis elements of $\na$ are
\bea
\nonumber \hat E_1 & = & 0,  \ \ \hat E_2= e_{1} \frac{\pd}{\pd e_n}, \ \ \hat E_k  = e_{k-2} \frac{\pd}{\pd e_{n-1}} +e_{k-1} \frac{\pd}{\pd e_n}, \; 3\leq k \leq n-2 ,\\
\hat E_{n-1}  & = &  - \sum_{k=3}^{n-2} e_{k-2} \frac{\pd}{\pd e_k}+ e_{n-2} \frac{\pd}{\pd e_n},\ \ \hat E_n  =  - \sum_{k=2}^{n-1} e_{k-1} \frac{\pd}{\pd e_k}. \label{dona}
\eea
The form of $\hat E_k, \ 1\leq k \leq n$ implies that the invariants do not depend on $e_{n-1},e_{n}$. Using Eq. (\ref{nomcas}) we find that the nilpotent Lie algebra $\na$ has $n-4$ functionally independent invariants
but it is rather complicated to directly solve the remaining two partial differential equations defining the invariants, namely 
$$\hat E_{n-1} \ I(e_1,e_2,\ldots,e_{n-2})=0, \qquad \hat E_{n} \ I(e_1,e_2,\ldots,e_{n-2})=0$$
(in particular, it is not too difficult to solve it for given $n=5,6,7,8,\ldots$ but it is more involved to deduce a general formula valid for arbitrary $n$ out of the solutions thus obtained). 

Let us employ the method of moving frames. Firstly, we construct the flows of the vector fields $\hat E_{n-1},\hat E_{n}$ acting on the space spanned by $e_1,\ldots,e_{n-2}$ only. We find
\begin{equation}
e_k(\Psi^{\alpha_{n-1}}_{\hat E_{n-1}} p) = \sum_{j=0}^{\left[ \frac{k-1}{2} \right]} \frac{(-1)^j}{j!} \alpha_{n-1}^j e_{k-2j}(p), \quad 1\leq k\leq n-2
\end{equation}
(where $[ \; \; ]$ denotes the integer part) and
\begin{equation}
e_k(\Psi^{\alpha_{n}}_{\hat E_{n}} p) = \sum_{j=0}^{k-1} \frac{(-1)^j}{j!} \alpha_{n}^j e_{k-j}(p), \quad 1\leq k\leq n-2.
\end{equation}
Combining these two expressions together we find (in the notation of the previous subsection)
\begin{equation}\label{tildee}
\tilde{e}_k = \sum_{l=0}^{k-1} \sum_{m=0}^{\left[ \frac{k-l-1}{2} \right]} \frac{(-1)^{l+m}}{l! m!} \alpha_{n}^l  \alpha_{n-1}^m e_{k-l-2m}, \quad 1\leq k \leq n-2. 
\end{equation}
We see that these $n-2$ functions involve only the group parameters $\alpha_{n-1},\alpha_{n}$. We can easily determine them out of two equations defining our section $\Sigma$. We choose them to be 
\begin{equation}\label{eqalpha}
0=\tilde{e}_2= e_2 - \alpha_{n} e_1, \qquad 0=\tilde{e}_3= e_3-\alpha_{n} e_2+\frac{\alpha_{n}^2}{2} e_1 - \alpha_{n-1} e_1.
\end{equation}
We find
\begin{equation}\label{solvalpha}
\alpha_{n-1} = \frac{1}{e_1^2}(e_1 e_3 -\frac{1}{2} e_2^2), \quad \alpha_{n} = \frac{e_2}{e_1}.
\end{equation}
Substituting these back into remaining equations (\ref{tildee}) and multiplying all of them by $e_1^{k-2}$ (in order to get polynomial expressions) we find the invariants
\begin{eqnarray}
 \xi_0 & = & e_1, \label{nilinv} \\
\nonumber \xi_j & = & e_1^{j+1} \tilde{e}_{j+3} = e_1^{j+1} \sum_{l=0}^{j+2} \sum_{m=0}^{\left[ \frac{j-l}{2}+1 \right]} \frac{(-1)^{l+m}}{l! m!} \alpha_{n}^l  \alpha_{n-1}^m e_{j+3-l-2m},   \quad 1\leq j \leq n-5. 
\end{eqnarray}
where the substitution (\ref{solvalpha}) is assumed. Performing it explicitly we arrive at the following theorem.
\begin{veta}\label{th2}
The nilpotent Lie algebra $\na$ has $n-4$ functionally independent Casimir invariants. They can be chosen to be the following polynomials
\bea
\nn \xi_0 & = & e_1, \\
\label{nac} \xi_j & = & \sum_{l=0}^{j+2} \sum_{m=0}^{\left[ \frac{j-l}{2}+1 \right]} \sum_{q=0}^{m} \frac{(-1)^{l+m+q}}{2^q \,l! (m-q)! q! } \, e_1^{j+1-l-m-q}\, e_2^{l+2q}\, e_3^{m-q}\, e_{j+3-l-2m}
\eea 
where $1\leq j \leq n-5$. All other Casimir invariants are functions of $\xi_0,\ldots,\xi_{n-5}$.
\end{veta}

We notice that $\xi_j $ is for $j\geq 1$ a homogeneous polynomial of degree $j+2$. For reader's convenience we list a few lowest order invariants explicitly (note that the dimension $n$ of $\na$ doesn't enter directly into the formulae, it just specifies where the list terminates)
\begin{eqnarray}
\nonumber \xi_1 & = & e_1^2 e_4-e_1 e_2 e_3 +\frac{1}{3} e_2^3 ,\\
\nonumber \xi_2 & = & e_1^3 e_5-e_1^2 e_2 e_4-\frac{1}{2} e_1^2 e_3^2+e_1 e_2^2 e_3-\frac{1}{4} e_2^4,\\
\xi_3 & = & e_1^4 e_6-e_2 e_1^3 e_5-e_1^3 e_3 e_4+e_2^2 e_1^2 e_4+ e_1^2 e_2 e_3^2- e_1 e_2^3 e_3+\frac{1}{5} e_2^5 ,\\
\nonumber \xi_4 & = & e_1^5 e_7- e_1^4 e_2 e_6+ e_1^3 e_2^2 e_5-e_1^4 e_3 e_5+ e_1^3 e_2 e_3 e_4-\frac{2}{3}  e_1^2 e_2^3 e_4 + \\
\nonumber & & + \frac{2}{3} e_1 e_2^4 e_3 + \frac{1}{3} e_1^3 e_3^3-e_1^2 e_2^2 e_3^2-\frac{1}{9} e_2^6 .
\end{eqnarray}

\subsection{The generalized Casimir invariants of the Lie algebra $\s_{n+1}$}

Let us now consider the $(n+1)$ dimensional solvable Lie algebra $\s_{n+1}$ of Theorem \ref{th1}. The operators $\hat E_i$ representing elements in the nilradical $\na$ will each contain an additional term involving a derivative with respect to $f_1$ and there is one additional operator, namely
\be
\label{F}
 \hat F_1 = -\sum_{k=1}^{n} (n-k+1) e_k \frac{\pd}{\pd e_k}.
\ee
However, from the form of these operators, namely $\hat E_1$, we see that the invariants cannot depend on $f_1$. Moreover, they can only depend on the invariants (\ref{nac}) of $\na$. To find the invariants of the algebra $\s_{n+1}$ we represent the non--nilpotent element $f_1 \in \s_{n+1}$ by the appropriate ``truncated'' differential operator acting only on $(e_1,\ldots,e_{n-2})$
\be\label{FT}
 \hat F_{1T} = -\sum_{k=1}^{n-2} (n-k+1) e_k \frac{\pd}{\pd e_k}.
\ee
We must then solve the equation \be\label{FTI}
\hat F_{1T} I(\xi_0,\ldots,\xi_{n-5})=0.
\ee
We can proceed directly, using an easily established formula
\begin{equation}\label{FTinv}
  \hat F_{1T} \, y(\xi_0,\ldots,\xi_{n-5})= n \xi_0 \frac{\partial y}{\partial \xi_0}+ \sum_{j=1}^{n-5} (j+2) (n-1) \xi_j \frac{\partial y}{\partial \xi_j}
\end{equation}
(in order to deduce Eq. (\ref{FTinv}) it is sufficient to determine how $\hat F_{1T}$ acts on each monomial in Eq. (\ref{nac})). 

Let us instead employ the method of moving frames, finding the flow of the vector field $F_{1T}$
\begin{equation}
e_k \left( \Psi^{\alpha_{n+1}}_{\hat F_{1T}} p \right) = \exp(-(n-k+1) \alpha_{n+1}) \, e_{k}(p), \quad 1\leq k\leq n-2.
\end{equation}
The full action of the group $S_{n+1}$ on the space with coordinates $e_1,\ldots,e_{n-2}$ gives
\begin{equation}\label{tildees}
\tilde{e}_k = \exp(-(n-k+1) \alpha_{n+1}) \sum_{l=0}^{k-1} \sum_{m=0}^{\left[ \frac{k-l-1}{2} \right]} \frac{(-1)^{l+m}}{l! m!} \alpha_{n}^l  \alpha_{n-1}^m e_{k-l-2m}, \quad 1\leq k \leq n-2. 
\end{equation}
We choose our section $\Sigma$ in the truncated space to be $\{ (1,0,0,\e_4,\ldots,e_{n-2}) \}$, i.e. we have one more equation in addition to Eq. (\ref{eqalpha})
\begin{equation}
 1=\tilde{e}_1= \exp(-n \alpha_{n+1}) e_1.
\end{equation}
Solving it we find 
\begin{equation}
\exp(- \alpha_{n+1}) = \left(\frac{1}{e_1}\right)^\frac{1}{n}
\end{equation}
and substituting it together with Eq. (\ref{solvalpha}) back into (\ref{tildees}) we find invariants which can be succinctly expressed in the form
\begin{equation}
\tilde{e}_k  =  \frac{\xi_{k-3}}{e_1^{\frac{jn+2n-j-2}{n}}}, \quad  4\leq k \leq n-2.
\end{equation}
By taking suitable powers (in order to express invariants as ratios of polynomials) we arrive at a theorem.
\begin{veta}\label{th5}
The $(n+1)$ dimensional solvable Lie algebra $\s_{n+1}$ of Theorem \ref{th1} has $n-5$ functionally independent invariants. They can be chosen in the form
\be\label{sinv}
\chi_j = \frac{\xi_j^{n}}{\xi_0^{(n-1)(j+2)}}, \ 1 \leq j \leq n-5, 
\ee
i.e. they are rational in $\xi_k$ and consequently in $e_k$.
\end{veta}

\section{Conclusions}

Let us sum up the main results of this paper and compare them with those obtained for indecomposable solvable Lie algebras $\s$ with other nilradicals of dimension $\dim \NR(\s)=n$.

\begin{enumerate}

\item The nilradical $\na$ (this article). The series exists for all $n\geq 5$. It is possible to add precisely one non--nilpotent element to form a solvable Lie algebra. Its action on the nilradical is a diagonal one. The algebra $\na$ has $n-4$ functionally independent Casimir operators. Its solvable extension has $n-5$ functionally independent generalized Casimir invariants, all of them can be chosen as rational functions of the elements of the nilradical.

\item The nilradical $\n_{n,1}$ (\cite{SW}). The series exists for all $n\geq 4$. It is possible to add at most $f_{\rm max}=2$ nil--independent element to $\n_{n,1}$. The algebra $\n_{n,1}$ has $n-2$ functionally independent Casimir invariants. For solvable Lie algebras we have $n-2-f$ independent generalized Casimir invariants. For $f=2$ they can be chosen to be rational functions, for $f=1$ they either can be chosen rational or they might involve logarithms. In both cases they depend only on the elements of the nilradical.

\item Abelian algebras $\a_n$ as nilradicals (\cite{NW,NW1,Ndo}). They exist for all $n\geq 1$. The Abelian algebra $\a_n$ has $n$ functionally independent Casimir operators (i.e. the basis elements of $\a_n$). For $n=1$ we can add just one non--nilpotent element. The obtained solvable Lie algebra has no generalized Casimir invariants. For $n=2$ over the field $F=\C$ we can add only one non--nilpotent element $f_1$ and we obtain a solvable algebra with one generalized Casimir invariant. Depending on the action of $f_1$ on the nilradical the invariant can be chosen rational or involves logarithms. For $n=2$ and $F=\R$ we can add one or two nil--independent elements. The obtained solvable algebras have just one independent invariant or none at all. For $n\geq 3$ we can add $f$ elements where $1 \leq f \leq n-1$. The number of independent generalized Casimir invariants of the solvable Lie algebras is $n-f$ and in general they may involve logarithms (i.e. cannot be expressed as rational). In all cases the generalized Casimir invariants depend on the elements of the nilradical alone. For $F=\R$ the generalized Casimir operators may involve other functions than logarithms, for instance inverse trigonometric ones. Implicit examples are in Ref. \cite{PSW} and explicit ones in Refs. \cite{BPP1,BPP2}.

\item Heisenberg algebras $\h(N)$ as nilradicals (\cite{RW}). The dimension of the Heisenberg algebra $\h(N)$ in $N$ dimensions is $n=2 N+1$ with $N \geq 1$. We can add up to $N+1$ nil--independent elements. The nilpotent algebra $\h(N)$ has only one Casimir invariant, corresponding to the one--dimensional center of $\h(N)$ spanned by $e_1$ (for any $N$). Two types of solvable extensions exist. If one of the non--nilpotent elements, say $f_1$, of the solvable Lie algebra does not commute with the center of $\h(N)$, i.e. with $e_1$, then all $f_a$'s must commute among each other 
$ [f_a,f_b]=0 $
and the solvable Lie algebra has $f-1$ independent generalized Casimir invariants. They can be chosen to be rational functions and depend both on elements of the nilradical $e_i$ and on the elements $f_a$. If we have
$$ [f_a,e_1]=0, \qquad 1\leq a \leq f$$
then the number of generalized Casimir invariants is $f+1-\gamma$ where $\gamma$ is the rank of the matrix $\gamma^1_{ab}$ in
\begin{equation}
[f_a,f_b]= \gamma^1_{ab} \, e_1.
\end{equation}
They can be chosen to be rational functions of elements $e_i$ and  $f_a$.

\item Triangular nilradicals $\t(N)$ (\cite{TW}). 
These can be represented by the set of all strictly upper triangular matrices. The dimension is $\dim \t(N)=n=\frac{N(N-1)}{2}$. For $N=3$ we have $\t(3)=\h(1)$, so the series really starts with $N=4$ and hence $n=6$. It is possible to add $f$ linearly independent elements to $\t(N)$ with $1\leq f \leq N-1$. The nilpotent algebra $\t(N)$ has $\left[ \frac{N}{2} \right] $ independent Casimir invariants. Let us denote by $S(N,f)$ the solvable Lie algebras that have $\t(N)$ as their nilradical and $f$ added non--nilpotent elements $f_a$. The Casimir invariants of $\t(N)$ and $S(4,f)$ with $f=1,2,3$ where calculated in Ref. \cite{TW1} using the infinitesimal method. Formulae for the case of $S(N,N-1)$ and $S(N,1)$ were also presented \cite{TW1} for all $N$ but the result was not proven and left as a conjecture. This result was later proven by Boyko et al. \cite{BPP3,BPP4} using the method of moving frames.
The algebras $S(N,N-1)$ have $\left[ \frac{N-1}{2} \right] $ independent generalized Casimir invariants. They can be chosen rational and involve both $e_i$ and $f_a$. The algebras $S(N,1)$ have either $\left[ \frac{N}{2} \right]+1 $ or $\left[ \frac{N}{2} \right]-1 $ generalized Casimir invariants; they can all be chosen rational and depend on $e_i$ and $f_1$.
Boyko et al. also calculated \cite{BPP5} the Casimir invariants for $S(N,f)$, $2 \leq f \leq N-2$ for the case when the non--nilpotent elements act diagonally. The invariants for a non--diagonal action of the nil--independent elements and even their number are known only for S(4,2) where there are either two invariants, or none \cite{TW1}. 

\end{enumerate}
This list of known series of solvable Lie algebras with given nilradicals covers with the notable exception of Abelian nilradicals only nilpotent algebras with one--dimensional center. This is not totally unexpected since larger center means that for given choice of derivations $D^1,\ldots,D^f$ there is more freedom in the choice of structure constants $\gamma^j_{ab}$ in Eq. (\ref{Agam2}) and consequently the classification becomes more intricate to perform and the results may be more complicated. It would be of interest to fully classify at least one such series.

As was mentioned in the Introduction such a list can never exhaust all possible solvable Lie algebras but it is possible to further extend it by considering other possible series of nilradicals. Such lists may be useful in testing some general hypotheses concerning solvable Lie algebras, e.g. the structure of their generalized Casimir invariants. 

\ack The research of Libor \v Snobl was supported by the research plan MSM6840770039 and the project LC527 of the Ministry of Education of the Czech Republic. The research of Pavel Winternitz was partly supported by a research grant from NSERC of Canada. He also thanks P. Exner and I. Jex for hospitality and the project LC06002 of the Ministry of Education of the Czech Republic for support during his visit to the Czech Technical University in Prague.

\end{document}